\documentclass[a4paper,fleqn]{cas-dc}

\usepackage[numbers]{natbib}

\def\tsc#1{\csdef{#1}{\textsc{\lowercase{#1}}\xspace}}
\tsc{WGM}
\tsc{QE}
\tsc{EP}
\tsc{PMS}
\tsc{BEC}
\tsc{DE}

\begin{document}
\let\WriteBookmarks\relax
\def\floatpagepagefraction{1}
\def\textpagefraction{.001}
\shorttitle{Coarse-to-fine Airway Segmentation Using Multi information Fusion Network and CNN-based Region Growing}
\shortauthors{J. Guo,R. Fu,B. He et~al.}

\title [mode = title]{Coarse-to-fine Airway Segmentation Using Multi information Fusion Network and CNN-based Region Growing}                      


%

\author[1]{Jinquan Guo}[type=editor,
                        ]

\credit{Conceptualization of this study, Methodology, Software}

\address[1]{School of Mechanical engineering and Automation, Fuzhou University, Fuzhou 350108, China}

\author[1]{Rongda Fu}[style=chinese]

\credit{Data curation, Writing - Original draft preparation}

\address[2]{School of Physics and Information Engineering, Fuzhou University, Fuzhou 350108, China}

\author[2]{Lin Pan}[style=chinese]

\author[2]{Shaohua Zheng}[style=chinese]

\author[2]{Liqin Huang}[style=chinese]

\author[3]{Bin Zheng}[style=chinese]
\cormark[1]
\ead{lacustrian@163.com}

\author[1]{Bingwei He}[auid=000,bioid=1,
                        orcid=0000-0001-7511-2910]
\cormark[2]
\ead{mebwhe@fzu.edu.cn}

\address[3]{Thoracic Department, Fujian Medical University Union Hospital.}

\cortext[cor1]{Corresponding author}
\cortext[cor2]{Principal corresponding author}


\begin{abstract}
Background and Objectives: Automatic airway segmentation from chest computed tomography (CT) scans plays an important role in pulmonary disease diagnosis and computer-assisted therapy. However, low contrast at peripheral branches and complex tree-like structures remain as two mainly challenges for airway segmentation. Recent research has illustrated that deep learning methods perform well in segmentation tasks. Motivated by these works, a coarse-to-fine segmentation framework is proposed to obtain a complete airway tree.

\noindent Methods: Our framework segments the overall airway and small branches via the multi-information fusion convolution neural network (Mif-CNN) and the CNN-based region growing, respectively. In Mif-CNN, atrous spatial pyramid pooling (ASPP) is integrated into a u-shaped network, and it can expend the receptive field and capture multi-scale information. Meanwhile, boundary and location information are incorporated into semantic information. These information are fused to help Mif-CNN utilize additional context knowledge and useful features. To improve the performance of the segmentation result, the CNN-based region growing method is designed to focus on obtaining small branches. A voxel classification network (VCN), which can entirely capture the rich information around each voxel, is applied to classify the voxels into airway and non-airway. In addition, a shape reconstruction method is used to refine the airway tree. 

\noindent Results: We evaluate our method on a private dataset and a public dataset from EXACT’09. Compared with the segmentation results from other methods, our method demonstrated promising accuracy in complete airway tree segmentation. In the private dataset, the Dice similarity coefficient (DSC), false positive rate (FPR), and sensitivity are 92.8\%, 0.015\%, and 88.6\%, respectively. In EXACT’09, the DSC, FPR, and sensitivity are 95.8\%, 0.053\% and 96.6\%, respectively.

\noindent Conclusion: The proposed Mif-CNN and CNN-based region growing method segment the airway tree accurately and efficiently in CT scans. Experimental results also demonstrate that the framework is ready for application in computer-aided diagnosis systems for lung disease and other related works.

\end{abstract}




\maketitle

\section{Introduction}

Chronic obstructive pulmonary disease (COPD) is the third leading cause of death, and it accounts for more than million deaths in China \cite{2018Prevalence}. Computed tomography (CT) technology is an important tool for the qualitative and quantitative assessment of lung tissue function, and it can improve the accuracy of diagnosis and treatment of pulmonary diseases (e.g., COPD). Airway segmentation from chest CT images can be used for many applications. First, it can help doctors make clinical decisions (e.g., disease diagnosis, surgical navigation, and evaluation of disease evolution). Second, the lungs are anatomically subdivided on the basis of the airway tree, and thus, the airway tree can facilitate the accurate definition of intersegmental demarcation, which is the most important step of thoracoscopic pulmonary segmentectomy. In addition, each airway branch is accompanied by an artery, and both structures have similar orientation; doctors also use the airways to distinguish the corresponding arteries in clinical practice \cite{Zhao2017B}. Experienced doctors manually label airway from CT via some interactive software, such as MIMICS and ITK-SNAP \cite{2009V}. However, manual segmentation is time consuming and susceptible to errors due to the large number of slices in CT images. Therefore, a robust automated airway segmentation method is necessary.

\begin{figure*}
\centering
\includegraphics[scale=.55]{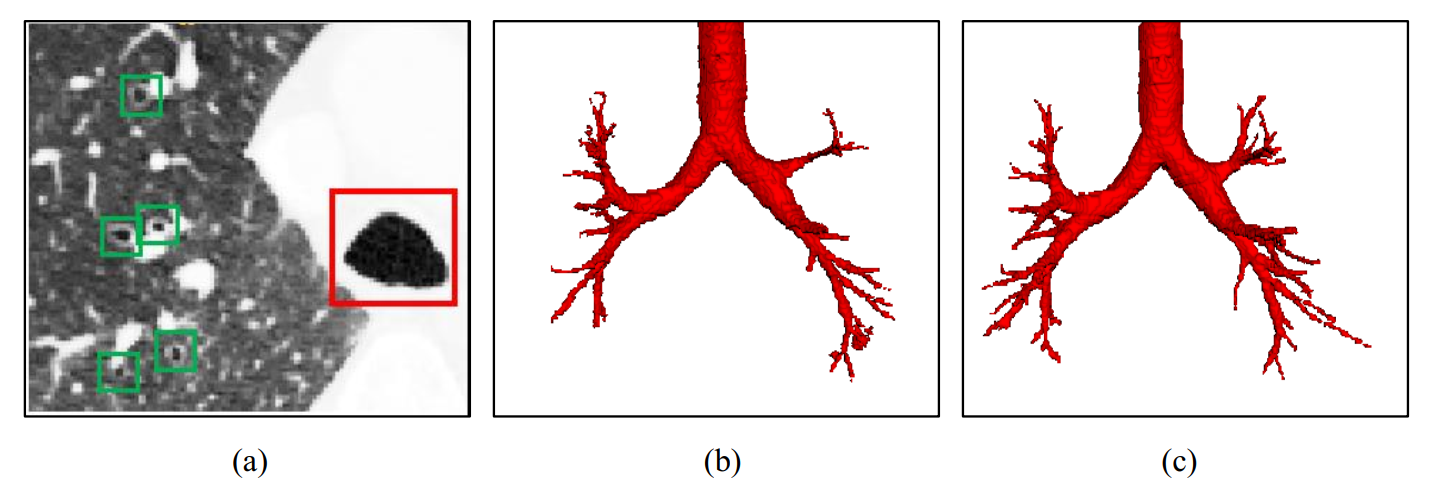}
\caption{(a) Axial view image showing the differences between large (red box) and small branches (green boxes), (b) incomplete airway tree, (c) accurate airway tree labeled by experienced radiologists}
\label{FIG:1}
\end{figure*}

Automatic airway segmentation suffers from several challenges. First, the size, shape and intensity of airway branches are various. Fig. \protect\ref{FIG:1} (a) illustrates the differences between large and small branches in axial view. The large one is easily identified, thus the lumen and lumen wall can be separated. However, the boundaries of small branches are blurred, and similar to the surrounding tissues. Second, the complex tree-like structure of the airway. Although airway branches have some common characteristics (e.g., direction and sequence order), the practical bronchi from different patients exhibit various appearances and shapes \cite{500140}. As a result, segmenting a complete airway tree becomes more difficult. Fig. \ref{FIG:1} (b) shows an incomplete airway tree obtained by U-Net \cite{2016U}. Fig. \ref{FIG:1} (c) shows an accurate airway manually labeled by experienced doctors. Compared with manually segmented trees, U-Net misses some important branches. 

Conventional airway segmentation methods generally include region-growing based methods, morphological methods and rule-based methods. A comparison of fifteen conventional airway segmentation methods was summarized in the EXACT’ 09 challenge in 2012 \cite{2012Exact09}. Conventional airway segmentation, which may also depend on the quality of the CT scan, is often a tedious task. Fine-tuning the parameters to achieve a balance between obtaining more airway branches and avoiding leakages is necessary in these methods \cite{Charbonnier2017}. Moreover, conventional airway segmentation methods rely on the expertise and experience of researchers to extract features.

Recently, deep learning technology has made remarkable improvements in the field of computer vision, and has become the most widely used approach in medical image segmentation. Qier et al. \cite{2017Tracking} proposed a method to segment the airway tree automatically by combining a fully-convolutional network (FCN) with image-based tracking algorithm. Charbonnier et al. \cite{Charbonnier2017} extracted the trachea and main bronchus by the region growing method, and a 2D convolution neural network (CNN) was used to segment small bronchi and remove leaks. Jin et al. \cite{20173D} proposed a 3D FCN to generate a probability map, and then a graph-based method which incorporates fuzzy connectedness segmentation was applied to refine the FCN output and guide leakage removal. Juarez et al. \cite{2018Automatic} proposed a method based on 3D U-Net. They also investigated the importance of data augmentation and loss function selection for airway segmentation. Qin et al. \cite{2019AirwayNet} provided a voxel-connectivity aware method, which focuses on reducing false positives and increasing the airway tree length. The author transformed a binary segmentation task into 26 tasks of predicting whether a voxel is connected to its neighbors. Zhao et al. \cite{Zhao2017B} used a two-stage 2D+3D neural network to segment thick and thin bronchi separately. Then the results from both stages was combined by a linear programming-based tracking algorithm. Qin et al. \cite{2020Learning} replaced the bottom layer of the U-Net with 3D slice-by-slice convolutional layers, which can capture the spatial information of elongated structures and improve the segmentation performance.

Although CNN-based methods are widely used for airway segmentation due to high sensitivity and less false negative rate, deep learning methods still suffer from some shortcomings. First, these methods integrate multi-scale contextual information via successive pooling and subsampling layers that reduce resolution until a global prediction is obtained \cite{Yu2016}. Segmenting small branches in the peripheral region is difficult. Second, some deep learning approaches are mainly based on intensity features and ignore other information (e.g., location information and boundary information), which can potentially improve the performance of airway segmentation.

In order to solve the aforementioned problems, we propose a segmentation framework which contains two parts: a multi-information fusion CNN (Mif-CNN) and a CNN-based region growing. The Mif-CNN is used to segment the overall airway. It combines multi-information, i.e., boundary and location information, to utilize additional feature knowledge. An atrous spatial pyramid pooling (ASPP) block is also integrated with Mif-CNN to obtain multiscale information. The CNN-based region growing method focuses on obtaining small branches. A voxel classification network (VCN) is applied to extract the airway voxel by voxel in the peripheral region.

Our main contributions are summarized as follows:
\begin{enumerate}[(1)]
\item We propose a coarse-to-fine segmentation framework to obtain airway tree. Our method focuses on improving the completeness of the airway tree.
\item We propose an Mif-CNN, which integrates ASPP, an edge guidance module (EGM) and the coordinates information of voxels with a u-shaped network. The network can utilize additional useful feature information to improve segmentation performance.
\item We propose a CNN-based region growing to segment the bronchi. The VCN can entirely capture the rich information around each voxel and facilitates the classification of voxels in peripheral regions into airway and non-airway voxels.
\end{enumerate}


\section{Method}

\begin{figure*}
\centering
\includegraphics[width=6.4in,height=3.2in]{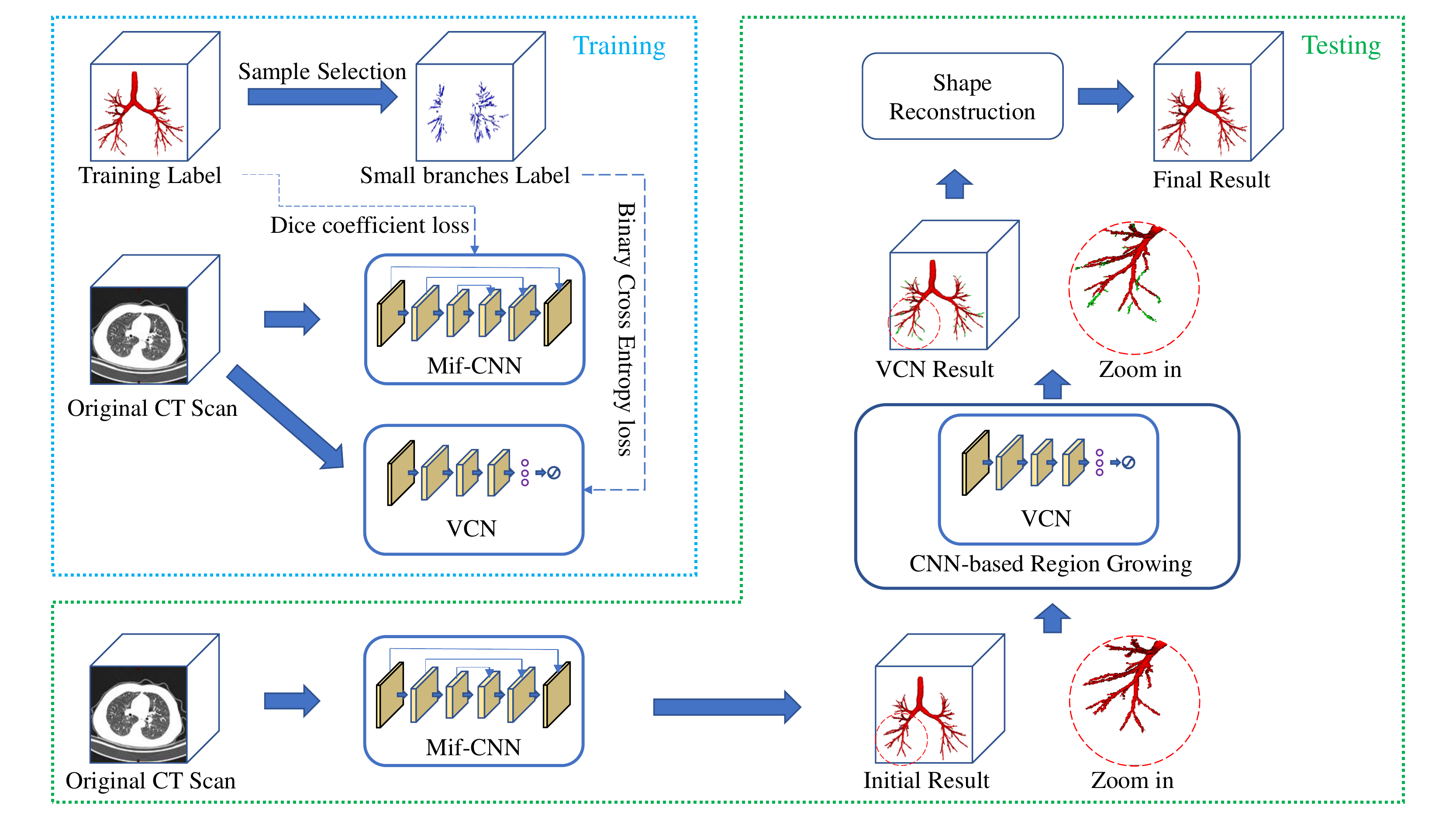}
\caption{Schematic of the workflow of our coarse-to-fine airway segmentation framework}
\label{FIG:2}
\end{figure*}

In this section, we introduce our method for airway segmentation. Section 2.1 describes the selection of training samples. Section 2.2 and Section 2.3 provides details of our Mif-CNN and CNN-based region growing method,respectively. Lastly, Section 2.4 presents our shape reconstruction method based on the centerline tracking algorithm.

 The workflow of our method is illustrated in Fig. \protect\ref{FIG:2}. For the training process, we first use sample selection to obtain small branches label, which contains subsegmental bronchi and segmental bronchi, whose diameters are less than 2 mm. Overall training label and small branches label are used to train the Mif-CNN and the VCN, respectively. The ASPP, EGM, and coordinate information are integrated with a u-shaped network in Mif-CNN; thus, additional context information and useful features are fused to improve segmentation performance. In the CNN-based region growing method, VCN is applied to classify airway voxels and non-airway voxels in the peripheral region.

For the testing process, the original CT scan is the only input, and selecting samples is unnecessary. We first obtain an initial airway tree by Mif-CNN. Then, we extract airway candidate voxels around the end of the initial tree. The VCN is used as a discriminator in the region growing method to classify voxels into airway or non-airway voxels. A voxel with high probability is considered an airway voxel and is connected to the airway tree, thus becoming the new endpoint of the tree. The iterative update stops until the airway tree is unchanged. Lastly, the result of VCN is refined by a shape reconstruction method.

\subsection{Training sample selection}

The coarse branches are easy to be extracted due to its large volume in airway. However, small branches are very thin and difficult to segment. We plan to learn the characteristics of the coarse branches and the small branches separately due to their differences in location, size, shape, intensity. Thus, the training data of Mif-CNN and VCN are different. Our sampling strategy follows:

\textbf{Dividing airway branches:} We obtain an airway skeleton tree with n branches \{$B_1$, $B_2$,…, $B_N$\} by an iterative backtracking algorithm presented in \cite{2018Rivulet}. Then we obtain the diameter $d_{i,j}$ of each voxel $v_{i,j}$,j$\in$ \{1,2,…,$K_i$\} in branch $B_i$. Thus, the branch diameter is computed as the average of all its centerline voxels’ diameters:

\begin{equation}
D_{B_i} = \frac{1}{K_i}\sum_{j=1}^{K_i}d_{i,j}
\end{equation}

Every branch is also labeled with corresponding anatomical name by a bronchus classification algorithm. Then all branches are classified into four types: trachea and main bronchus, lobar bronchi, segmental bronchi and subsegmental bronchi. We further sample two subsets from these branches in line with their branch measurements and anatomical level:

(a). The first subset $S_1$ contains whole airway tree;
(b). The second subset $S_2$ contains segmental bronchi whose diameter is less than 2 mm and subsegmental bronchi. 

\textbf{Sampling for the Mif-CNN:} We then obtain volume of interests (VOIs) from the first subsets$S_1$ to train Mif-CNN. Considering the limitation of computer GPU memory, we then apply overlapped sliding windows with VOIs sized to 64 × 64 × 64 pixels and the stride size is 32 during training and testing. 

\textbf{Sampling for the VCN:} Our voxel classification network is trained to classify a voxel point into airway or non-airway in the peripheral region. Therefore, we select sample points from $S_2$, and obtained VOIs from these voxel points.

We random select 3000 points from the $S_2$ in each CT scans as positive samples. Then we select non-airway voxels around the selected airway voxels. We divide non-airway points into five subsets by measuring Euclidean distance. In the five subsets, the distance between non-airway voxel and airway voxels is 1 voxel, 2–4 voxels, 5–7 voxels, 8–10 voxels, and 11–30 voxels, respectively. We random selected 600 points in every non-airway subset and obtain 3000 non-airway points as negative samples in total. For each sample points, we extract an VOIs sized to 32 × 32 × 32 pixels with this point as center.

\subsection{ Mif-CNN}

\begin{figure*}
\centering
\includegraphics[width=6in,height=2.8in]{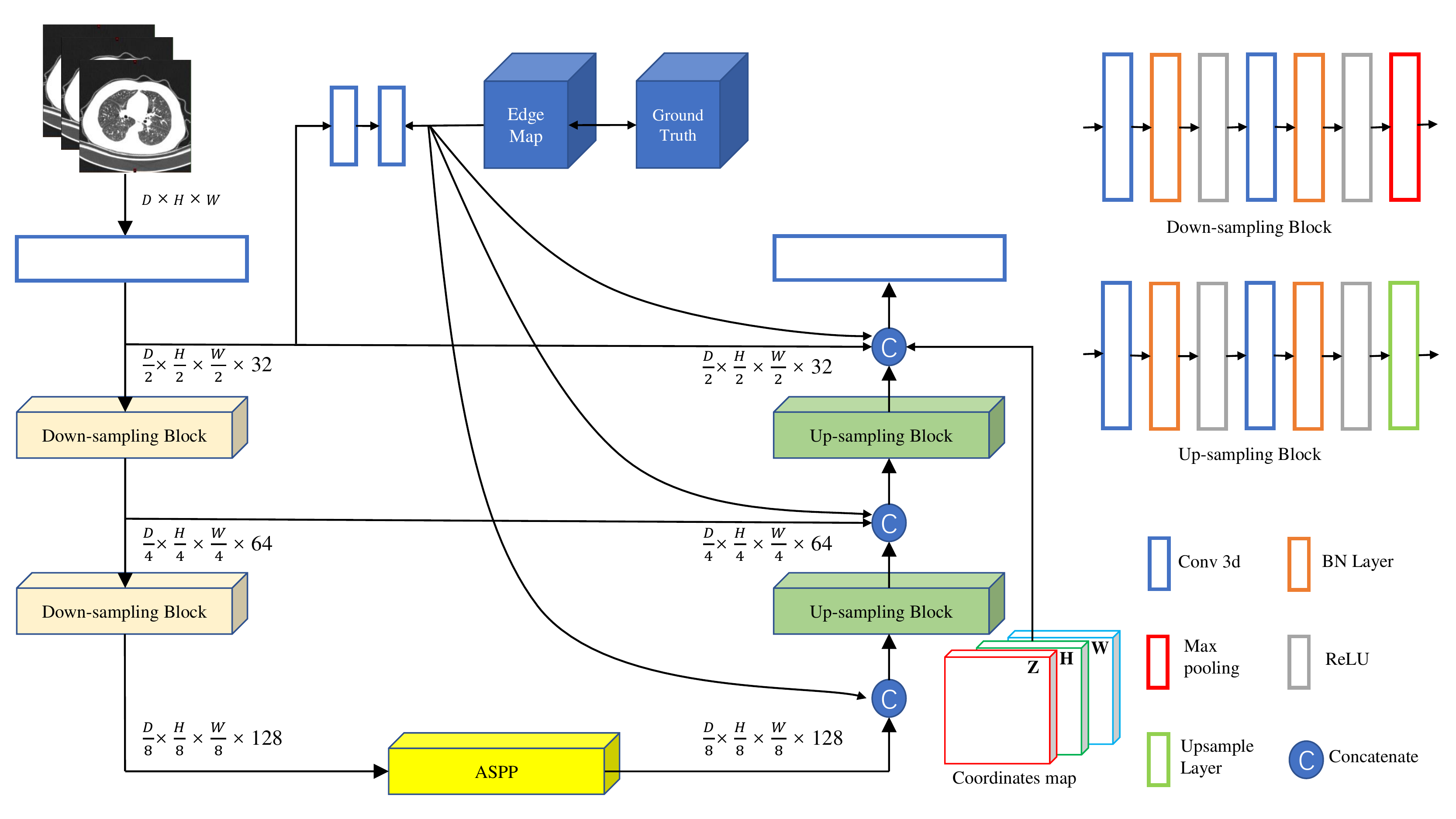}
\caption{The basic network of Mif-CNN}
\label{FIG:4}
\end{figure*}

The architecture of Mif-CNN is shown in Fig. \protect\ref{FIG:4}. The encoder-decoder structure needs to expand receptive field via successive pooling and subsampling layers, and it gradually reduces the spatial resolution. Features of thin bronchi, whose diameters are usually only 2-3 voxels, are prone to vanish after three times of pooling, making it difficult to segment and recover. In order to achieve the purpose of maintaining image resolution, our Mif-CNN only contains two pooling layers. However, multiple pooling layers are necessary to extract effective context information for large bronchi. Our method inspired with dilated/atrous convolution which can increase receptive field and maintain the number of kernel parameters at the same time. We combine the u-shape structure and dilated convolution to solve the problem of reducing the size of feature map and detect multi-scale object. For the purpose of improving the segmentation performance, we use boundary information and coordinate information to integrate more useful features. 

\begin{figure}[h]
\centering
\includegraphics[scale=.25]{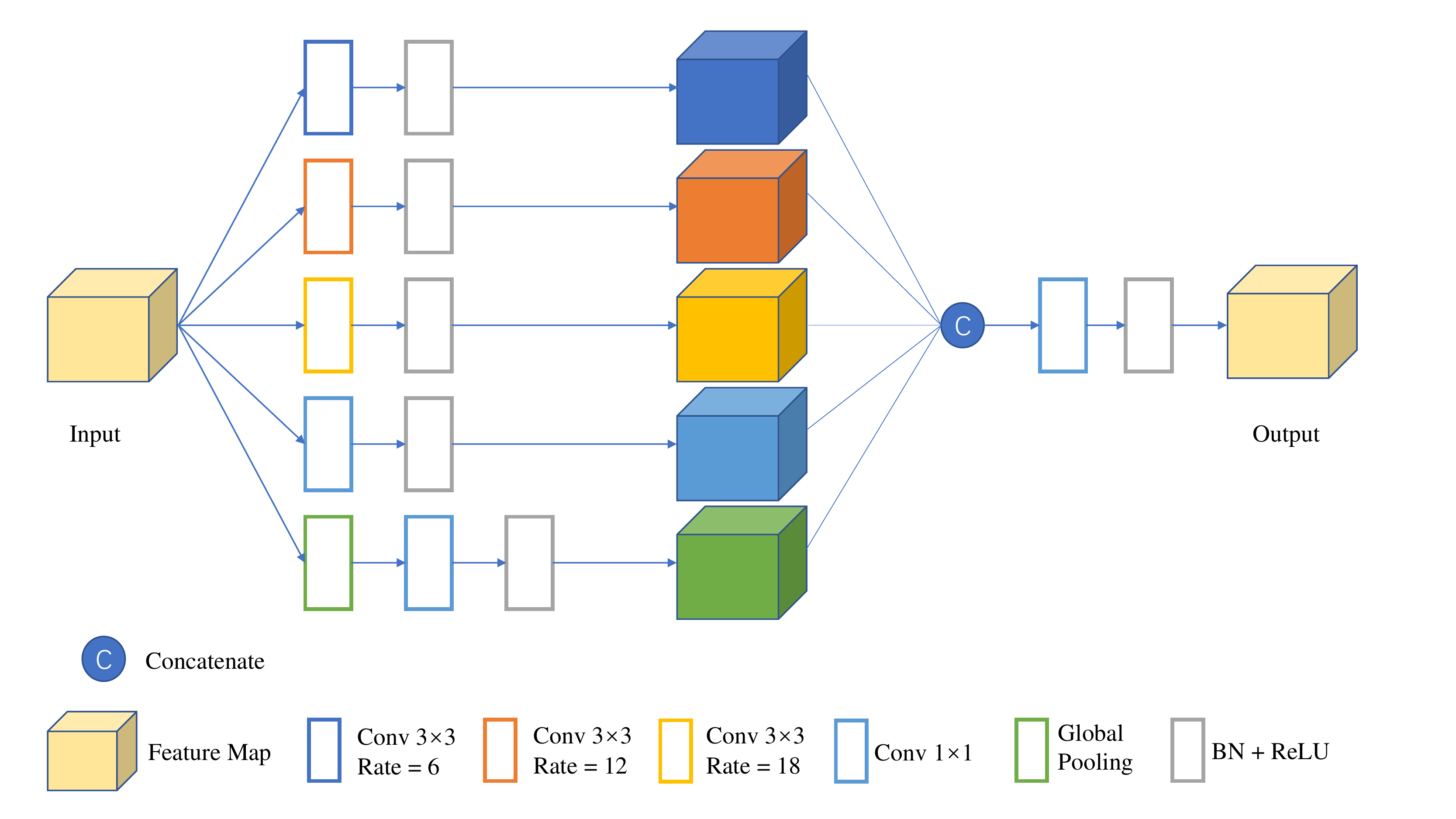}
\caption{The basic architecture of ASPP}
\label{FIG:5}
\end{figure}

\textbf{Atrous spatial pyramid pooling:} ASPP is useful to expand the receive field and resample features at multi-scale \cite{2017Aspp}. In order to capture context infromation at multi-range, we adopt a ASPP module as the bottom layer in the encoding-path of the network. As shown in Fig. \protect\ref{FIG:5}, the structure of ASPP module is mainly composed of two parts: (a) one 1×1 convolution and three 3 × 3 convolutions with dilated rates of 6, 12 and -18, (b) a global average pooling layer, and a 1 × 1 convolution. The first part is aim to expand the receive field and better obtain multi-scale feature maps, the second part is used to overcome the problem of effective weight reduction at long range. Moreover, all the convolution operations with batch normalization (BN) layers and rectified linear unit (ReLU). Finally, these feature maps generated from the five branches are concatenated, and sent to a 1 × 1 convolution with BN.

\textbf{Edge guidance module:} CNN method are mainly based on intensity features and ignore other information. Boundary information are essential characteristics for segmenting target in medical images, and it can improve feature detection for segmentation \cite{2019EGNet}. Therefore, we consider that low-level features are rich in spatial details, and can provide sufficient boundary information. We first detect local boundary information from low-level feature map, then propagate the location information of this layer from top to bottom. Moreover, the feature map is also fed to two successive convolution layers, and supervised by targeting on an edge map $M_e$ derived from ground truth masks. 

\textbf{Coordinate information:} We also consider that location information of voxels beneficial for the network to segment airway. Therefore, we convert the coordinate information of the voxel in three dimensions into a three-channel feature map consistent with the network size. Then concatenate them with the feature map on the last layer in decoder part of Mif-CNN.

\textbf{Loss function:} As shown in Fig. \protect\ref{FIG:4}, Mif-CNN has two head branches. We use a Binary Cross Entropy (BCE) loss function for supervising the feature map of EGM, which is formulated as:
\begin{equation}
\Gamma_{\text {edge }}=\sum_{x \in M_{e}} \log \left(1-p_{e}(x)\right)+\sum_{x \notin M_{e}} \log \left(p_{e}(x)\right)
\end{equation}

where x denote the voxel and $p_e$ (x) is the predict edge voxel.

The second branch is trained to predict whether each pixel is an airway voxel. The loss function used for our pixel level classification problem is Dice coefficient loss function:
\begin{equation}
\Gamma_{\text {seg }}=1- \sum_{i=1}^{2} \frac{2 * \left(p_{i} * g_{i}\right)}{p_{i}+g_{i}+\epsilon}
\end{equation}

Where $p_i$, and $g_i$,i$\in$\{0,1\} denote the predict segmentation result and label, respectively. $\epsilon$ is a smooth term to avoid division by zero.

Finally, we define our loss function $\Gamma_{total}$ as a combination of $\Gamma_{seg}$ and $\Gamma_{edge}$, which can be expressed by the following formula:
\begin{equation}
\Gamma_{\text {total}}=\Gamma_{\text {seg}}+\Gamma_{\text {edge}}
\end{equation}

\subsection{CNN-based region growing}

Conventional region growing methods are used to segment the airway based on their density (in Hounsfield Units or HU). However, since the density of airway voxels are close to the surrounding tissue at peripheral branches, these methods perform worse. In order to extract more airway voxel, we propose a CNN-based region growing method which combines deep learning technology and region growing. In this work, we first initialize seed point. The voxels in the terminal of the airway tree which obtained from Mif-CNN are selected as the initial seed points, and they are pushed on a stack. Unlike the work in \cite{2009Region}, we classify those unprocessed voxels in the 26-neighbor of the seed point into airway voxels and non-airway voxels by using a 3D CNN (VCN). VCN can produce a probability whether the voxels are airway voxels. A threshold $T_u$  ($T_u$=0.8) is introduced to discriminate airways and non-airways. A voxel with probability which higher than $T_u$ is considered as airway voxel, it will be selected as the new seed and pushed on the stack. The airway tree is allowed to keep growing when the stack is not empty.

\begin{figure}[h]
\centering
\includegraphics[scale=.25]{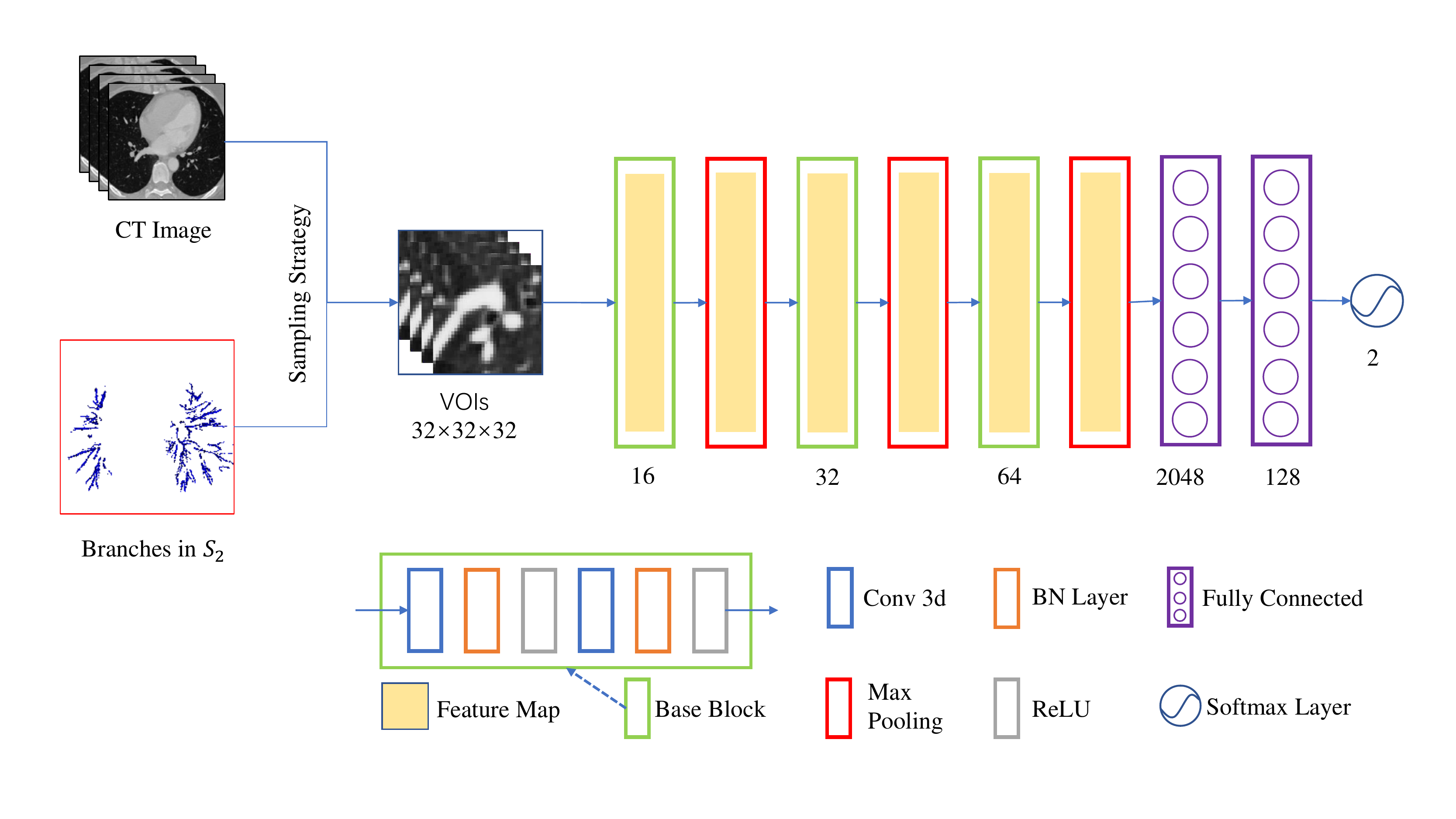}
\caption{Scheme of the VCN architecture. $S_2$ contains segmental bronchi whose diameter is less than 2 mm and subsegmental bronchi. Using VOIs of the voxels from $S_2$ as the input of VCN.}
\label{FIG:6}
\end{figure}

The architecture of VCN is shown in Fig. \protect\ref{FIG:6}. It consists of three down-sampling blocks and two fully connected layers. The down-sampling operation is used for encode high semantic feature. Down-sampling block consists of two 3 × 3 convolution layers with BN and ReLU, and each block followed by a 2 × 2 max pooling operation with stride 2. The number of kernels in each convolution layer is 16,16,32,32,64,64. After down sampling, we then feed the feature map into two successive fully connected layers which have 2048 neurons and 128 neurons respectively. Finally, we calculate the probabilities of airway voxel by a softmax function after the last fully connected layer. We use a Binary Cross Entropy (BCE) loss function for supervising the output. The loss is illustrated as:
\begin{equation}
\Gamma_{s}=-y \log p-(1-y) \log (1-p)
\end{equation}

where  y$\in$\{0,1\} is the predict probability and p$\in$[0,1] is the label.

\subsection{Shape reconstruction}

Our segmentation framework is robust due to its high quality of output and less false negative rate. Mostly, our method will not produce large leakages, but cause small leakages near the peripheral bronchi. Some small boundary protrusions, branch like structures adjacent to airway branches also exist. Hence, our shape reconstruction method based on centerline tracking algorithm is proposed to remove these leakages. The detailed description is as follows: 

\begin{enumerate}[(1)]
\item \textbf{Skeleton detection:} We use an iterative backtracking algorithm presented in \cite{2018Rivulet} to compute the skeleton of airway obtained by our network. Then we remove some wrong branches and construct an airway skeleton tree.
\item \textbf{Airway reconstruction:} For each voxel $v_{i,j}$, we calculate the average diameter of itself and four neighbors by the following formula: $d_{i, j}^{\prime}=\frac{1}{5} \sum_{k=-2}^{2} d_{i, j+k}$, we then replace $d_{i,j}$ as the new diameter of $v_{i,j}$. Then we reconstruct the airway via compute the ball with diameter $d_{i,j}$ of each voxel.
\item \textbf{Result refinement:} The reconstructed airway is used to refine the segmentation results. We supplement the small fractures which $d_{i, j}<\frac{2}{3} d_{i j}^{\prime}$, and voxel point outside the radius of reconstructed airway are removed. 
\end{enumerate}

\section{Experiments and Result}

In this section, several experiments are conducted to evaluate the Mif-CNN and VCN segmentation methods. In Section 3.1, information on our private data and the public data of EXACT’09 are presented.  In Section 3.2, we introduce evaluation metrics and experimental settings. In Sections 3.3 and 3.4, we show the segmentation results of Mif-CNN and the CNN-based region growing method, respectively. Given that our study focuses on extracting small branches in the peripheral region, the overall airway tree and the airway tree with removed trachea and main bronchus are evaluated.

\subsection{Dataset}

We evaluated our airway segmentation method on two datasets: private chest CT scans, and public CT scans of EXACT’ 09 \cite{2012Exact09}.

\textbf{Private dataset:} It consists of 20 CT scans. Each slice of the CT scans has the same size of 512 × 512 pixels with a spatial resolution ranging from 0.625 mm to 1 mm. The slice thickness varies from 0.5 mm to 1.25 mm, and the number of slices in each CT scan ranges from 237 to 441. On the basis of the interactive segmentation results of ITK-SNAP, the ground truths are manually corrected by two experienced doctors \cite{2009V}.

\textbf{EXACT’09:} It consists of 20 CT scans from the training dataset of this challenge. All slices in the CT have a size of 512 × 512 pixels, with a pixel size ranging from 0.5 mm to 0.78 mm in axial view. The number of slices in each CT scan ranges from 157 to 764, and the slice thickness varies from 0.45 mm to 1.0 mm. We obtain the annotation of 20 public CT scans from Qin \cite{2020AirwaynetSE}. The acquisition and investigation of data conform to the principles outlined in the Declaration of Helsinki \cite{2001World}.

\subsection{Evaluation metrics and experimental setting}

To evaluate the performance of our methods, four metrics based on area overlap are used: dice similarity coefficient (DSC), false positive rate (FPR), sensitivity (Sen) and precision (Pre). These metrics are illustrated as follows:
\begin{equation}
D S C=2 \times \frac{A_{p r e} \cap A_{g t}}{A_{p r e}+A_{g t}}
\end{equation}
\begin{equation}
F P R=2 \times \frac{F_{P}}{T_{P}+T_{P}}
\end{equation}
\begin{equation}
\text {Sen}=2 \times \frac{T_{P}}{T_{P}+F_{P}}
\end{equation}
\begin{equation}
\text {Pre}=2 \times \frac{T_{P}}{T_{P}+F_{N}}
\end{equation}

where $A_{pre}$ and $A_{gt}$ denote the segmentation result and ground-truth respectively. $T_P$ denote voxels of predicted positive class belong to true airway. $F_P$ denote voxels of predicted positive class belong to non-airway, $F_N$ denote voxels of predicted negative class belong to non-airway, $T_N$ denote voxels of predicted positive class belong to true airway.

We also removed trachea and main bronchus from both predictions and ground truth, while calculated these metrics.

In our experiments, we randomly chose 12 cases from private dataset and 12 cases from public dataset for training. Then we chose 3 cases from private dataset and public dataset respectively as validation dataset. The remaining 10 cases are test dataset. We implemented our method in PyTorch and fine-tuned the hyper-parameter on training data.

For the Mif-CNN, the Adam optimizer($\beta_1$ = 0.9, $\beta_2$ = 0.999) is used with a learning rate of 1.0e-04 in training. The maximum train epoch is set to 200. For the Voxel classification network, the SGD is used with a learning rate of 1.0e-04 in training. We finally use the model with best validation results for testing.

We then evaluating the performance of our Mif-CNN in our private dataset and public dataset of EXACT’09, respectively.

\subsection{Segmentation results of Mif-CNN}

To verify the segmentation performance of our Mif-CNN, we compare it with three state-of-the-art methods, i.e., those of Jin \cite{20173D}, and Juarez \cite{2018Automatic} and 3D U-Net\cite{2016U}. We reimplement these methods in PyTorch and fine-tune the hyperparameter. Detailed experiment results are shown as follows:

\subsubsection{Experiment on private dataset}

\begin{table}[width=1\linewidth,cols=5,pos=h]
\caption{Results of the proposed Mif-CNN in comparison with state-of-the-art methods (mean ± standard deviation) on the private testing dataset while evaluating the whole airway.}\label{tbl1}
\begin{tabular*}{\tblwidth}{@{} LLLLL@{} }
\toprule
Method & DSC (\%) & FPR (\%) & Sen (\%) & Pre (\%) \\
\midrule
U-Net \cite{2016U} & 91.6±1.6 & 0.018±0.007 & 88.7±3.6 & 94.8±1.7  \\
Jin \cite{20173D} & 92.4±2.3 & 0.022±0.010 & 88.8±4.9 & 95.7±1.5 \\
Juarez \cite{2018Automatic} & 92.8±2.6 & 0.015±0.007 & 88.6±4.8 & 97.4±1.3\\
Mif-CNN & 93.5±2.4 & 0.020±0.009 & 90.8±5.2 & 96.7±1.4 \\
\bottomrule
\end{tabular*}
\end{table}

\begin{table}[width=1\linewidth,cols=5,pos=h]
\caption{Results of the proposed Mif-CNN in comparison with state-of-the-art methods (mean ± standard deviation) on the private testing dataset while evaluating an airway without trachea and main bronchus.}\label{tbl2}
\begin{tabular*}{\tblwidth}{@{} LLLLL@{} }
\toprule
Method & DSC (\%) & FPR (\%) & Sen (\%) & Pre (\%) \\
\midrule
U-Net \cite{2016U} & 81.3±5.6 & 0.017±0.007 & 74.2±8.6 & 87.7±2.7\\
Jin \cite{20173D} & 82.1±4.9 & 0.016±0.006  & 76.5±8.8  & 91.7±3.2 \\
Juarez \cite{2018Automatic} & 83.4±5.6 & 0.012±0.008 & 75.1±9.1 & 92.8±2.2 \\
Mif-CNN & 84.6±5.7 & 0.018±0.009 & 78.8±9.6 & 92.3±2.4 \\
\bottomrule
\end{tabular*}
\end{table}

The result of the four segmentation methods for the overall airway tree are presented in Table \ref{tbl1}. On average, our method achieves the highest DSC of 93.5\%, which is 0.07\% higher than second segmentation performance of Juarez \cite{2018Automatic}. In terms of sensitivity, our method gets 2.0\% improvement than second performance of Jin \cite{20173D}. The experiment results illustrate that our method outperforms the other methods in detecting a complete airway tree.

The result of four segmentation methods for the airway tree without trachea and main bronchus are presented in Table \ref{tbl2}. By comparing U-Net, the DSC and sensitivity increase from 81.3\% and 74.2\% to 84.6\% and 78.8\%, respectively. In terms of FPR and precision, our method achieves a result of 0.02\% and 96.7\%. It could also shows that our method outperforms the other methods in the peripheral region.

\begin{figure*}
\centering
\includegraphics[scale=.58]{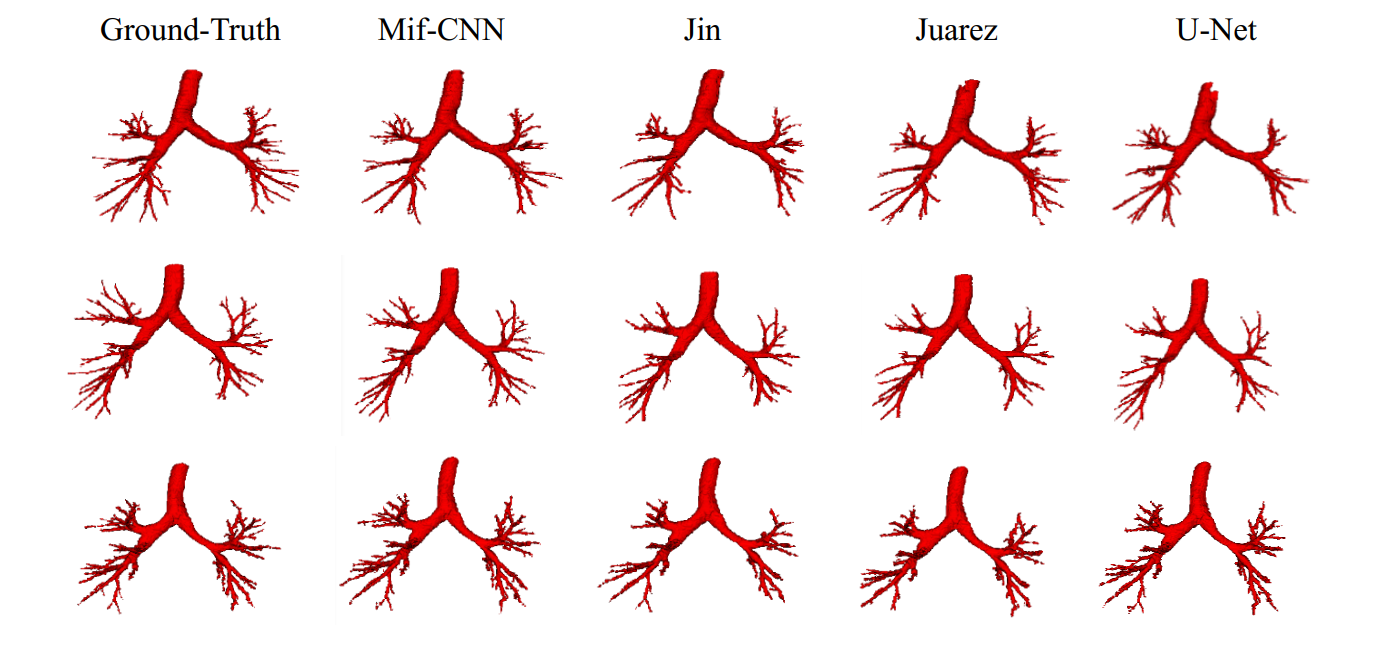}
\caption{Comparison of airway segmentation results between other methods and ground truth. From Row 1 to Row 3 are three different subjects. From left to right are the results of ground-truth, Mif-CNN, Jin, Juarez and U-Net respectively.}
\label{FIG:7}
\end{figure*}

\begin{figure*}
\centering
\includegraphics[scale=.58]{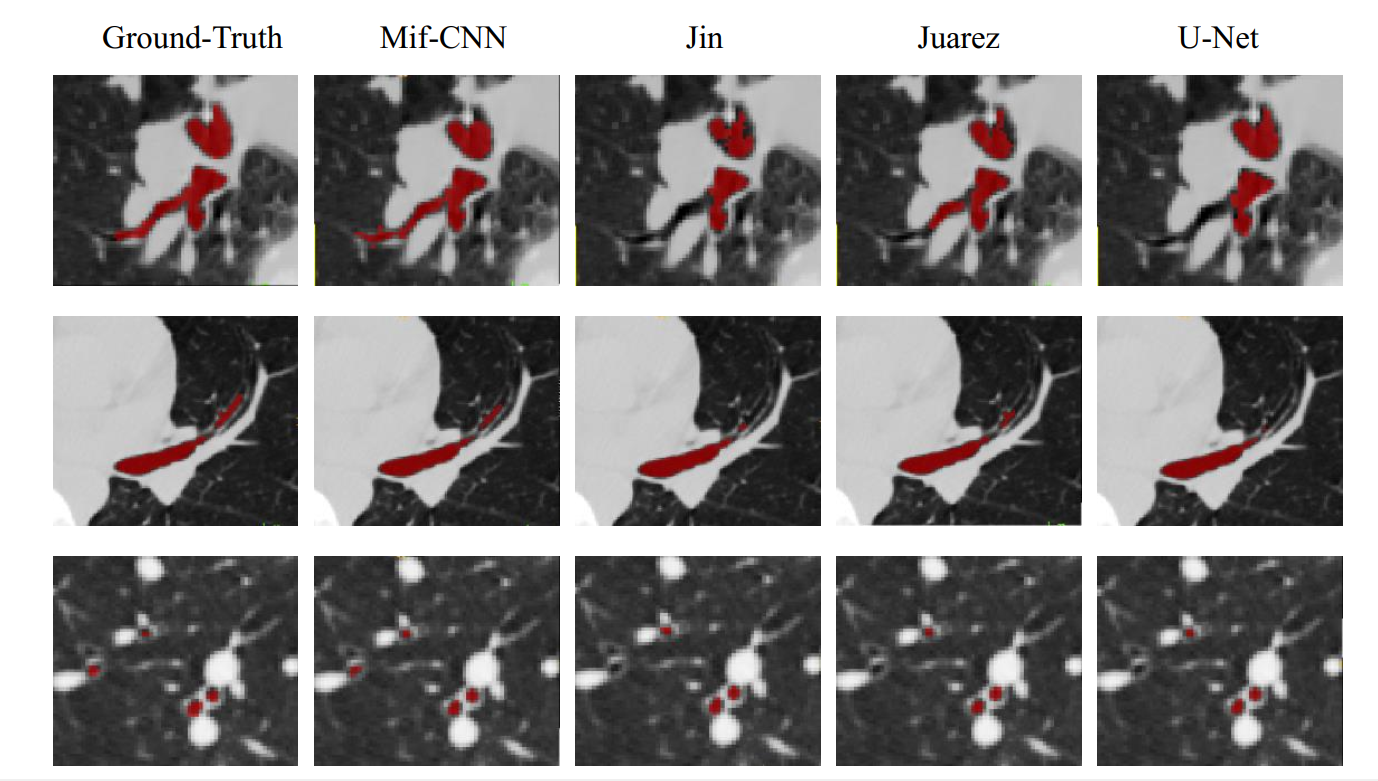}
\caption{Axial view images of three different subjects from Row 1 to Row 3. From left to right are the results of ground-truth, Mif-CNN, Jin, Juarez and U-Net respectively}
\label{FIG:8}
\end{figure*}

Fig. \ref{FIG:7} shows the segmentation results of three different subjects on 3D images. Compared with Jin ($3^rd$ column), Juarez ($4^th$ column) and U-Net ($5^th$ column), our Mif-CNN ($2^nd$ column) extracts more branches in the upper left lobe and upper right lobe, which are prone to lesions and nodules. A better airway segmentation result will facilitate the screening and diagnosis of lesions. Specifically, the result of our method is close to the ground-truth. In contrast, U-Net misses a lot of important branches in some case. Jin \cite{20173D} and Juarez \cite{2018Automatic} improve the segmentation results, but the performance is still leaving a little to be desired on detecting more peripheral branches.

Axial view of three subject for qualitative analysis are shown in Fig. \ref{FIG:8}. From the $1^st$ row, our method ($2^nd$ column) not loses important branches. It can be seen from the $2^nd$ row and $3^rd$ row, our method compared with other methods, still has a good performance in the low-contrast peripheral area, where it is difficult to distinguish the airway and surrounding tissues.

\subsubsection{Experiment on EXACT’09}

\begin{table}[width=1\linewidth,cols=5,pos=h]
\caption{Results of the proposed Mif-CNN in comparison with state-of-the-art methods (mean ± standard deviation) on the public testing dataset while evaluating the whole airway.}\label{tbl3}
\begin{tabular*}{\tblwidth}{@{} LLLLL@{} }
\toprule
Method & DSC (\%) & FPR (\%) & Sen (\%) & Pre (\%) \\
\midrule
U-Net \cite{2016U} & $95.7 \pm 1.1$ & $0.051 \pm 0.017$ & $96.4 \pm 1.8$ & $95.3 \pm 2.4$  \\
Jin \cite{20173D} & $95.7 \pm 1.1$ & $0.051 \pm 0.017$  & $96.4 \pm 1.8$ & $95.3 \pm 2.4$\\
Juarez \cite{2018Automatic}  & $95.8 \pm 1.1$ & $0.052 \pm 0.018$  & $96.5 \pm 1.9$ & $95.1 \pm 2.6$\\
Mif-CNN  & $95.8 \pm 1.3$ & $0.053 \pm 0.019$ & $96.6 \pm 1.5$ & $95.0 \pm 2.6$\\
\bottomrule
\end{tabular*}
\end{table}

\begin{table}[width=1\linewidth,cols=5,pos=h]
\caption{Results of the proposed Mif-CNN in comparison with state-of-the-art methods (mean ± standard deviation) on the public testing dataset while evaluating airway without trachea and main bronchus.}\label{tbl4}
\begin{tabular*}{\tblwidth}{@{} LLLLL@{} }
\toprule
Method & DSC (\%) & FPR (\%) & Sen (\%) & Pre (\%) \\
\midrule
U-Net \cite{2016U} & $83.2 \pm 3.5$ & $0.032 \pm 0.021$ & $83.7 \pm 5.6$ & $82.8 \pm 6.8$ \\
Jin \cite{20173D} & $84.1 \pm 4.3$ & $0.030 \pm 0.016$  & $85.1 \pm 7.9$ & $83.9 \pm 6.0$\\
Juarez \cite{2018Automatic} & $83.8 \pm 3.8$ & $0.031 \pm 0.018$ & $85.0 \pm 7.8$ & $83.6 \pm 6.3$ \\
Mif-CNN & $85.2 \pm 4.4$ & $0.033 \pm 0.018$ & $86.2 \pm 6.1$ & $82.6 \pm 6.9$ \\
\bottomrule
\end{tabular*}
\end{table}

For the public dataset, comparison results of proposed method and the state-of-the-art methods are illustrated in Table \ref{tbl3}. Our Mif-CNN achieves 95.8\%, 0.053\%, 95.0\% and 96.6\% for DSC, FPR, precision and sensitivity, respectively. Compared to the results by U-Net \cite{2016U}, the scores of DSC and sensitivity are 0.1\% and 0.2\% higher. The overall performance of our method is better.

We further evaluate the performance of all methods in the peripheral region. The results are shown in Table \ref{tbl4}. Our Mif-CNN achieves 85.2\%, 0.033\%, 82.6\% and 86.2\% for DSC, FPR, precision and sensitivity, respectively. In terms of FRP, Jin \cite{20173D} achieves the best result of 0.03\%, but the score for DSC and sensitivity are 1.1\% and 1.2\% lower than our method, respectively. Considering that our purpose is to extract more airway, the false positive rate does not affect the overall result. Therefore, these results verify the feasibility and effectiveness of Mif-CNN.

\subsection{Segmentation results of the CNN-based region growting}

\begin{figure*}
\centering
\includegraphics[scale=.4]{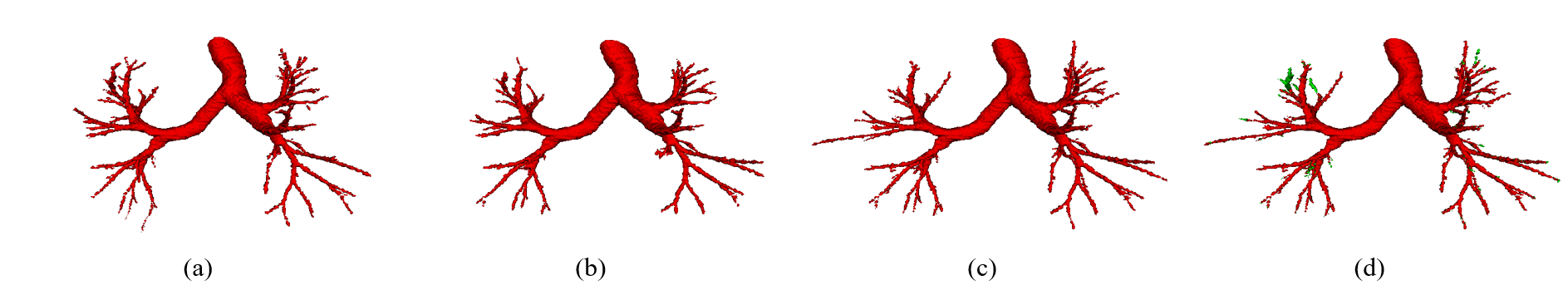}
\caption{Refined segmentation results using our voxel classification network. (a) Gold standard labeled by an experimental radiologist, (b) Result of U-Net, (c) Initial output of Mif-CNN, (d) Final result of the voxel classification network.}
\label{FIG:9}
\end{figure*}

Although the Mif-CNN (Fig. \ref{FIG:9} (c)) can extract more airway branches than U-Net (Fig. \ref{FIG:9} (b)), it still misses several branches. Our CNN-based region growing method focuses on classifying airway voxels at peripheral branches, and the VCN is applied to produce the probability of whether a voxel belongs to the airway. Therefore, we extracted airway-candidate VOIs from the sample points located in 26-neighbor voxels from the end points of the initial airway, and classified them as airway or non-airway by VCN. The airway result was updated by using only the voxels with high probability ($\geq$ 0.8), and connected it to the initial airway. The updated airway was input to the extraction process of the airway-candidate VOIs of the next iteration. The proposed CNN-based region growing method iteratively updates the initial airway result until the airway is unchanged after two successive updates.

We segment more airway voxels by using this method and obtain a more complete airway tree, which is close to the golden standard. An example of segmentation result is shown in Fig. \ref{FIG:9}. As shown in the figure, our CNN-based region growing method can extract some missing branches (represented in green color) in the peripheral region.  

\subsection{Ablation study}

\begin{table*}[width=1.5\linewidth,cols=5,pos=h]
\caption{Ablation study of the proposed Mif-CNN.}\label{tbl5}
\begin{tabular*}{\tblwidth}{@{} LLLLL@{} }
\toprule
Method & DSC (\%) & FPR (\%) & Sen (\%) & Pre (\%) \\
\midrule
(No. 1) U-Net & 81.3 & 0.017 & 74.2 & 93.1 \\
(No. 2) U-Net + CI & 82.2 & 0.015 & 74.9 & 93.6 \\
(No. 3) U-Net + EGM & 82.1 & 0.020 & 76.4 & 94.2 \\
(No. 4) U-Net + ASPP & 82.2 & 0.024 & 78.3 & 91.6 \\
(No. 5) U-Net + CI + ASPP & 82.9 & 0.021 & 77.3 & 93.2 \\
(No. 6) U-Net + CI + EGM & 82.6 & 0.018 & 76.8 & 95.1 \\
(No. 7) U-Net + CI + ASPP + EGM & 84.6 & 0.016 & 78.8 & 92.3 \\
\bottomrule
\end{tabular*}
\end{table*}

Our Mif-CNN is mainly based on integrating multiple basic modules and prior information. Thus, the effectiveness of the key components of our network must be studied through detailed ablation experiments. We evaluate three key components of our network: the coordinate information, the ASPP, and the EGM. The ablation study results are shown in Table \ref{tbl5}. The trachea and main bronchus are not included in this evaluation.

\begin{enumerate}[(1)]
\item \textbf{Effectiveness of Coordinate Information:} Theoretically, location information facilitates the object targeting. Thus, we add coordinate information to our network and further investigate its roles in our network. The results (No.2 vs. No.1, No.5 vs. No.3, No.6 vs. No.4) show that the models with coordinate information have a higher precision and lower FPR than those without coordinate information. The coordinate information restricts the spatial position of the airway and avoids leakage of the peripheral bronchi. However, the models with coordinate information have lower sensitivity than those that without.
\item \textbf{Effectiveness of ASPP:} We also explore the contribution of the ASPP module. Given the atrous convolution that expands the receptive field without reducing spatial resolution, our network can maximize multi-information and has an improved ability to detect small objects. As shown in Table \ref{tbl5}, No. 4 performs better than other settings (Nos.1, 2, and 3) and achieves the highest sensitivity at 78.3\%. This result indicates that the ASPP module, which can resample features at multi-scales, can obtain more airway voxels. Although the ASPP module results in a slightly higher FPR, this FPR does not have much effect on our results. The extraction of additional peripheral branches proves that our experiment is successful.
\item \textbf{Effectiveness of EGM:} Boundary information is essential for segmentation targets in medical images, and it can improve edge feature detection for segmentation. We use an EGM to propagate rich boundary information from low-level features to a high-level feature map. In this manner, boundary information can improve segmentation performance. Compared with modules without edge constraint guidance (No. 1), our method is improved by 2.2\% in sensitivity and by 1.1\% in precision.
\end{enumerate}

\section{Discussion}

In the experiments, we evaluate the performance of the proposed Mif-CNN and VCN. In Section 3.3.1, we first investigate Mif-CNN compared with other methods, including that of Jin and Juarez and U-Net, on the private dataset. The results indicate that our method achieves a better performance on the private dataset than on the public one. The DSC and sensitivity for the whole airway segmentation are 93.5\% and 90.8\%, respectively. The result is promising as compared with the result from 3D U-Net, which is 91.6\% in DSC and 88.7\% in sensitivity. The DSC and sensitivity of the quantitative results without the trachea and main bronchus also show that Mif-CNN has an improved performance in airway segmentation. In addition, the qualitative analysis illustrates that the segmentation result by Mif-CNN is well, which is close to the ground truth.

In Section 3.3.2, we compare the performance of Mif-CNN and other methods on the public dataset of EXACT’09. As shown in Table \ref{tbl3}, our method performs well but does not show a significant difference compared with other methods. Nevertheless, our method outperforms others when the trachea and main bronchus are removed from the airway tree (Table \ref{tbl4}). The reason is that lobar bronchi and segmental bronchi only account for approximately 30\% of the total airway volume. When the segmentation result achieves a slight improvement in the peripheral region, the numerical change when computing the metrics of the whole airway tree is not significantly different.

In Section 3.4, we study the improvement of segmentation results by the CNN-based region growing method. The initial airway tree of Mif-CNN is better than U-Net and is close to the ground truth. However, it also misses several branches. Fig. \ref{FIG:9} shows the result of the CNN-based region growing method performs better than the initial airway tree, indicating that the voxel-by-voxel classification approach based on the deep neural network is useful. VCN can entirely capture rich information around each voxel and improve the segmentation performance. 

Our study focuses on extracting more airway branches in the peripheral region. Thus, we also discuss the results of two different datasets, which evaluate the airway tree without trachea and main bronchus. As shown in Table \ref{tbl2} and \ref{tbl4}, our method achieves the best result in peripheral region regardless of which dataset is used. We attribute this result to the EGM and coordinate information, which fuse additional explicit knowledge to produce useful features. In addition, the ASPP can expand the receptive field and detect multi-scale information, thus improving the segmentation performance. False positives are higher than some methods because some of these errors might be due to miss terminal branches in the reference standard; however, our method can successfully extract these missing peripheral branches. In addition, these real branches are considered false positive branches when calculating the FPR metrics. The multiple-dataset validation also demonstrates that our method is robust and reliable by fusing additional useful information.

In the literature, a few studies have focused on automatic airway segmentation. One of the classic methods is that of \cite{2010Vessel}, in which a vessel-guided airway segmentation algorithm that extracts additional airway voxels by combining vessel information was proposed; the TPR reached up to 98.68\%. A conference paper \cite{2019AirwayNet} also leveraged distance to the lung border and voxels’ coordinates to improve airway segmentation. This operation integrated additional semantic information to improve learning of additional airway features, and the DSC was 90.2\% on 10 chest CT scans. These studies believe that the fusion of additional feature information can improve segmentation performance. Thus, we use an EGM to integrate additional edge information. The author in \cite{2018Small} focused on the small branches. A random forest classifier was applied to classify each voxel into airway voxel or non-airway voxel. In \cite{YUN201913}, a 2.5D CNN was used to classify each voxel at the peripheral branches. Three patches for axial, coronal, and sagittal views in each voxel are fed into the network, and the CNN was then used for voxel classification. The DSC of this method could reach up to 89.97\%. However, the data and implementation of different studies vary, thus, comparing different quantitative results objectively is difficult.

\section{Conclusion}

This paper has presented a fully automatic method to perform segmentation of airways from chest CT scans. The core of this method is a coarse-to-fine framework which segments the overall airway tree and small airway branches respectively. The framework contains two parts: Mif-CNN and CNN-based region growing. In Mif-CNN, ASPP, EGM and coordinate information are incorporated into a u-shape network, thus it can utilize more useful features to improve the performance of segmentation. CNN-based region growing method can extract branches which are missing in the result of Mif-CNN. In addition, a shape reconstruction method based on centerline tracking algorithm is employed to refine the final segmentation result. Experimental results on private dataset and public dataset show that this method performance well in the upper lobe region which prone to lesions and nodules. The multiple-dataset validation also demonstrates the reliability and further practicality of the proposed method. In the future, we plan to apply our segmentation results to other related tasks, such as Artery/Vein separation, determination of lung segments. 

\section*{Acknowledgement}
This work was supported by the Natural Science Foundation of Fujian Province, China (Grant No.2020J01472) and Provincial Science and Technology Leading Project (Grant No.2018Y0032).

\bibliographystyle{unsrt}   

\bibliography{cas-refs}

\begin{thebibliography}{10}

\bibitem{2018Prevalence}
Chen Wang, Jianying Xu, and et~al. Lan~Yang.
\newblock Prevalence and risk factors of chronic obstructive pulmonary disease
  in china (the china pulmonary health [cph] study): a national cross-sectional
  study.
\newblock {\em The Lancet}, 391(10131), 2018.

\bibitem{Zhao2017B}
Tianyi Zhao, Zhaozheng Yin, Jiao Wang, Dashan Gao, Yunqiang Chen, and Yunxiang.
  Mao.
\newblock Bronchus segmentation and classification by neural networks and
  linear programming.
\newblock {\em Medical Image Computing and Computer Assisted Intervention --
  MICCAI 2019}, pages 230--239, 2019.

\bibitem{2009V}
E.M Van~Rikxoort, W~Baggerman, and B~van Ginneken.
\newblock Automatic segmentation of the airway tree from thoracic ct scans
  using a multi-threshold approach.
\newblock {\em Second International Workshop on Pulmonary Image Analysis},
  486:341–349, 2009.

\bibitem{500140}
M.~{Sonka}, {Wonkyu Park}, and E.~A. {Hoffman}.
\newblock Rule-based detection of intrathoracic airway trees.
\newblock {\em IEEE Transactions on Medical Imaging}, 15(3):314--326, 1996.

\bibitem{2016U}
Çiçek Ö., Ahmed Abdulkadir, Soeren~S. Lienkamp, Thomas Brox, and Olaf.
  Ronneberger.
\newblock 3d u-net: Learning dense volumetric segmentation from sparse
  annotation.
\newblock {\em Medical Image Computing and Computer-Assisted Intervention --
  MICCAI 2016}, 9901:424--432, 2016.

\bibitem{2012Exact09}
P.~{Lo}, B.~{van Ginneken}, J.~M. {Reinhardt}, T.~{Yavarna}, P.~A. {de Jong},
  B.~{Irving}, C.~{Fetita}, M.~{Ortner}, R.~{Pinho}, J.~{Sijbers},
  M.~{Feuerstein}, A.~{Fabijanska}, C.~{Bauer}, R.~{Beichel}, C.~S. {Mendoza},
  R.~{Wiemker}, J.~{Lee}, A.~P. {Reeves}, S.~{Born}, O.~{Weinheimer}, E.~M.
  {van Rikxoort}, J.~{Tschirren}, K.~{Mori}, B.~{Odry}, D.~P. {Naidich},
  I.~{Hartmann}, E.~A. {Hoffman}, M.~{Prokop}, J.~H. {Pedersen}, and M.~{de
  Bruijne}.
\newblock Extraction of airways from ct (exact'09).
\newblock {\em IEEE Transactions on Medical Imaging}, 31(11):2093--2107, 2012.

\bibitem{Charbonnier2017}
Jean~Paul Charbonnier, Eva~M.van Rikxoort, Arnaud~A.A. Setio, Cornelia~M.
  Schaefer-Prokop, Bram van Ginneken, and Francesco Ciompi.
\newblock Improving airway segmentation in computed tomography using leak
  detection with convolutional networks.
\newblock {\em Medical Image Analysis}, 36:52--60, 2017.

\bibitem{2017Tracking}
Qier Meng, Holger~R. Roth, Takayuki Kitasaka, Masahiro Oda, Junji Ueno, and
  Kensaku Mori.
\newblock Tracking and segmentation of the airways in chest ct using a fully
  convolutional network.
\newblock {\em Medical Image Computing and Computer Assisted Intervention --
  MICCAI 2017}, 10434:198--207, 2017.

\bibitem{20173D}
Dakai Jin, Ziyue Xu, Adam~P. Harrison, Kevin George, and Daniel~J. Mollura.
\newblock 3d convolutional neural networks with graph refinement for airway
  segmentation using incomplete data labels.
\newblock {\em International Workshop on Machine Learning in Medical Imaging},
  10541:141--149, 2017.

\bibitem{2018Automatic}
Antonio Garcia-Uceda~Juarez, H.~A. W.~M. Tiddens, and M.~de~Bruijne.
\newblock Automatic airway segmentation in chest ct using convolutional neural
  networks.
\newblock {\em Image Analysis for Moving Organ, Breast, and Thoracic Images},
  11040:238--250, 2018.

\bibitem{2019AirwayNet}
Yulei Qin, Mingjian Chen, Hao Zheng, Yun Gu, Mali Shen, Jie Yang, Xiaolin
  Huang, Yue-Min Zhu, and Guang-Zhong Yang.
\newblock Airwaynet: A voxel-connectivity aware approach for accurate airway
  segmentation using convolutional neural networks.
\newblock {\em Medical Image Computing and Computer Assisted Intervention --
  MICCAI 2019}, 11769:212--220, 2019.

\bibitem{2020Learning}
Yulei Qin, Hao Zheng, Yun Gu, Xiaolin Huang, Jie Yang, Lihui Wang, and Yue-Min
  Zhu.
\newblock Learning bronchiole-sensitive airway segmentation cnns by feature
  recalibration and attention distillation.
\newblock {\em Medical Image Computing and Computer Assisted Intervention --
  MICCAI 2020}, 12261:221--231, 2020.

\bibitem{Yu2016}
Fisher Yu and Vladlen Koltun.
\newblock Multi-scale context aggregation by dilated convolutions.
\newblock {\em 4th International Conference on Learning Representations, ICLR
  2016 - Conference Track Proceedings}, 2016.

\bibitem{2018Rivulet}
S.~{Liu}, D.~{Zhang}, Y.~{Song}, H.~{Peng}, and W.~{Cai}.
\newblock Automated 3d neuron tracing with precise branch erasing and
  confidence controlled back tracking.
\newblock {\em IEEE Transactions on Medical Imaging}, 37(11):2441--2452, 2018.

\bibitem{2017Aspp}
Liang{-}Chieh Chen, George Papandreou, Florian Schroff, and Hartwig Adam.
\newblock Rethinking atrous convolution for semantic image segmentation.
\newblock {\em CoRR}, abs/1706.05587, 2017.

\bibitem{2019EGNet}
J.~{Zhao}, J.~{Liu}, D.~{Fan}, Y.~{Cao}, J.~{Yang}, and M.~{Cheng}.
\newblock Egnet: Edge guidance network for salient object detection.
\newblock pages 8778--8787, 2019.

\bibitem{2009Region}
Eva van Rikxoort, W.~Baggerman, and B.~Ginneken.
\newblock Automatic segmentation of the airway tree from thoracic ct scans
  using a multi-threshold approach.
\newblock {\em The Second Workshop on Pulmonary Image Analysis}, pages
  341--349, 2009.

\bibitem{2020AirwaynetSE}
Y.~{Qin}, Y.~{Gu}, H.~{Zheng}, M.~{Chen}, J.~{Yang}, and Y.~{Zhu}.
\newblock Airwaynet-se: A simple-yet-effective approach to improve airway
  segmentation using context scale fusion.
\newblock pages 809--813, 2020.

\bibitem{2001World}
Article~In Italian.
\newblock World medical association (amm). helsinki declaration. ethical
  principles for medical research involving human subjects.
\newblock {\em Assistenza Infermieristica E Ricerca Air}, 20(2):104, 2001.

\bibitem{2010Vessel}
Pechin Lo, Jon Sporring, Haseem Ashraf, Jesper~J.H. Pedersen, and Marleen {de
  Bruijne}.
\newblock Vessel-guided airway tree segmentation: A voxel classification
  approach.
\newblock {\em Medical Image Analysis}, 14(4):527 -- 538, 2010.

\bibitem{2018Small}
Z.~Bian, J~P Charbonnier, J.~Liu, D.~Zhao, D~A Lynch, and B~Van Ginneken.
\newblock Small airway segmentation in thoracic computed tomography scans: a
  machine learning approach.
\newblock {\em Physics in medicine and biology}, 63(15), 2018.

\bibitem{YUN201913}
Jihye Yun, Jinkon Park, Donghoon Yu, Jaeyoun Yi, Minho Lee, Hee~Jun Park,
  June-Goo Lee, Joon~Beom Seo, and Namkug Kim.
\newblock Improvement of fully automated airway segmentation on volumetric
  computed tomographic images using a 2.5 dimensional convolutional neural net.
\newblock {\em Medical Image Analysis}, 51:13 -- 20, 2019.

\end{thebibliography}



\end{document}